\documentclass[fleqn,10pt,onecolumn]{wlscirep}

\usepackage[utf8]{inputenc}
\usepackage[T1]{fontenc}

\usepackage{graphicx}
\usepackage[T1]{fontenc}
\usepackage[utf8]{inputenc}
\usepackage{authblk}
\usepackage{amsmath}
\usepackage{amssymb}
\usepackage{bm}
\usepackage{todonotes}
\usepackage{ulem} % allows strikethrough via \sout
\usepackage{indentfirst}
\usepackage{lineno}
\usepackage{verbatim}
\usepackage{placeins}
\newtheorem{defn}{Definition}
\newtheorem{rem}[defn]{Remark}

\providecommand{\R}{\ensuremath \mathbb{R}}

\newcommand{\regtext}[1]{\mathrm{\textnormal{#1}}}
\newcommand{\defemph}[1]{\emph{#1}}

\newcommand{\ff}{\regtext{ff}}
\newcommand{\lb}{^\regtext{lb}}
\newcommand{\ub}{^\regtext{ub}}
\newcommand{\upt}{^{(t)}}
\newcommand{\uptdt}{^{(t + \Delta t)}}

\title{Characterizing the limits of human stability during motion: perturbative experiment validates a model-based approach for the Sit-to-Stand task}
\author[1*]{Patrick D. Holmes}
\author[1]{Shannon M. Danforth}
\author[1,2]{Xiao-Yu Fu}
\author[3,4,5]{Talia Y. Moore}
\author[1,3]{Ram Vasudevan}
\affil[1]{Department of Mechanical Engineering, University of Michigan}
\affil[2]{Faculty of Kinesiology, University of Calgary}
\affil[3]{Robotics Institute, University of Michigan}
\affil[4]{Department of Ecology and Evolutionary Biology, University of Michigan}
\affil[5]{Museum of Zoology, University of Michigan}
\affil[*]{Send correspondence to pdholmes@umich.edu}
\begin{abstract}
Falls affect a growing number of the population each year.
Clinical methods to identify those at greatest risk for falls usually evaluate individuals while they perform specific motions such as balancing or Sit-to-Stand (STS).
Unfortunately these techniques have been shown to have poor predictive power and are unable to identify the magnitude, direction, and timing of perturbations that can cause an individual to lose stability during motion.
To address this limitation, the recently proposed Stability Basin (SB) aims to characterize the set of perturbations that will cause an individual to fall under a specific motor control strategy.
The SB is defined as the set of configurations that do not lead to failure for an individual under their chosen control strategy.
This paper presents a novel method to compute the SB and the first experimental validation of the SB with an $11$-person perturbative STS experiment involving forwards or backwards pulls from a motor-driven cable.
The individually-constructed SBs are used to identify when a trial fails, i.e., when an individual must switch control strategies (indicated by a step or sit) to recover from a perturbation.
The constructed SBs correctly predict the outcome of trials where failure was observed with over $90\%$ accuracy, and correctly predict the outcome of successful trials with over $95\%$ accuracy.
The SB was compared to three other methods and was found to estimate the stable region with over $45\%$ more accuracy in all cases.
This study demonstrates that SBs offer a novel model-based approach for quantifying stability during motion, which could be used in physical therapy for individuals at risk of falling.
\end{abstract}

\begin{document}
\flushbottom
\maketitle
\thispagestyle{empty}
\noindent \textbf{Keywords: biomechanics, mathematical model, locomotion, fall risk, reachability, feedback control}

\section{Introduction} 
Falls are the leading cause of injury in people over 75 years old, resulting in reduced quality of life, increased healthcare costs, and accident-related death \cite{rubenstein2006, stevens2006, heinrich2010}.
If at-risk individuals can be identified prior to injury, the likelihood of falling can be significantly reduced through intervention such as physical therapy \cite{robertson2001, bloem2001}.
Fall risk generally results from instability arising from neuromuscular deficiencies or external perturbations. 
Clinical assessments for identifying individuals who would benefit from preventative care are currently limited to questionnaires 
\cite{delbaere2009, hauer2009} and non-perturbative motor assessments \cite{berg1989, podsiadlo1991, lundin1997}. 
Self-reported information often has low reliability \cite{weinstein1987, furnham1986}, and current clinical motor performance tests have low fall-prediction rates, especially for active older adults \cite{shumway2000, steffen2002, laessoe2007, muir2008}.
For widespread use, stability assessments must combine predictive power with minimal experimental and computation time.

Since older adults are more likely to fall while in motion \cite{hausdorff2001}, several studies suggest that quantifying dynamic stability may help identify biomechanical deficiencies associated with an increased risk of falling \cite{lockhart2008, hsiao2008, hur2010, bruijn2014}.
Thus, a number of model-based methods have been developed to assess stability during walking \cite{bruijn2014}.
Among these, variability measures \cite{terrier2011} and the maximum Lyapunov exponent \cite{dingwell2000} ranked highest overall in validity.
However, these metrics only characterize a subject's ability to recover from small perturbations.
Currently, it is unclear whether the most useful estimates of stability are provided by measuring an individual's ability to recover from small perturbations or by computing the largest possible perturbation that they can withstand without switching control strategies or falling \cite{bruijn2014}.
Unfortunately, these methods cannot yet be fully compared since a verified technique for computing the maximum perturbation a subject can recover from does not exist.
Additionally, the best-performing model-based methods are limited to periodic motion.

Stability during non-rhythmic motion is of interest, especially since difficulty with aperiodic tasks such as Sit-to-Stand (STS) are strongly correlated with falls in older adults and necessary for maintaining independence and quality of life \cite{campbell1989}.
The motion of a person's center of mass (COM) during STS can be modelled as an inverted pendulum, which requires an appropriate amount of angular momentum at seatoff to successfully complete the task \cite{hof2005}.
Drawing from this idea, metrics of stability for STS are generally based on an individual's initial COM velocity or acceleration \cite{pai1997, simoneau2005, fujimoto2012}.
However, a stability metric that considers data only at the onset of movement disregards valuable information about the control strategy used by the individual.
In fact, human control strategies for STS lie on a spectrum ranging from Quasi-Static (QS), in which little momentum is used and the body position is statically stable throughout the motion, to Momentum-Transfer (MT), which is statically unstable and %\talia{\sout{makes extensive use of} 
extensively uses momentum to achieve standing \cite{hughes1994}.

To account for the dynamic differences in STS motions under distinct control strategies, Shia et. al introduced the Stability Basin (SB), computed using individualized pendulum models of STS with linear-quadratic regulator (LQR) controllers.
For this pendulum model, the SB at a given time is equal to the set of pendulum states that are able to successfully achieve standing under the specified controller.
Though this SB under an LQR controller successfully distinguishes between less and more stable STS strategies, its ability to identify exactly the set of states that can arrive at standing without switching strategies is unverified.

To test whether the SB can accurately estimate an individual's stability for a particular task, perturbations must be introduced in a way that causes the states of the individualized dynamic model to exit the stable region.
An accurate prediction of instability corresponds to an experimentally observed failure of the individual's control strategy.

Here we use an $11$-person perturbative STS experiment, described in Sec. \ref{sec:methods:experimental_protocol}, to validate stability predictions generated by the SB method.
Study participants attempted to complete the STS motion with their natural, QS, and MT control strategies while subjected to forwards or backwards cable pulls, applied to their approximate COM.
We use a dynamic model, presented in Sec. \ref{sec:methods:subsec2}, and subject-specific controllers to form individualized SBs, then test whether each subject's SB can accurately predict a switch in control strategies in order to recover from the cable pull, described in Sec. \ref{sec:methods:subsec4}.
We consider common failure modes, such as a step or sit, to be observed changes in STS control strategy \cite{riley1998}.
In contrast to prior work that assumed that humans use LQR control during STS while constructing the SB, we use a data-driven method to create bounds on the control input for a particular subject and control strategy, described in Sec. \ref{sec:methods:subsec3}.
Across all subjects and control strategies, the proposed method correctly predicts unsuccessful trials over $90\%$ of the time, and correctly identifies successful trials over $95\%$ of the time in a leave-one-out assessment.
When compared to three other methods of assessing STS stability, the SB method out-performed each alternative.
This perturbative validation experiment demonstrates that the SB method is accurate and reliable, showing promise for integration into current healthcare infrastructure.

\section{Methods} \label{sec:methods}

\begin{figure}[!b]
	\centering
	\includegraphics[width = 0.5\linewidth]{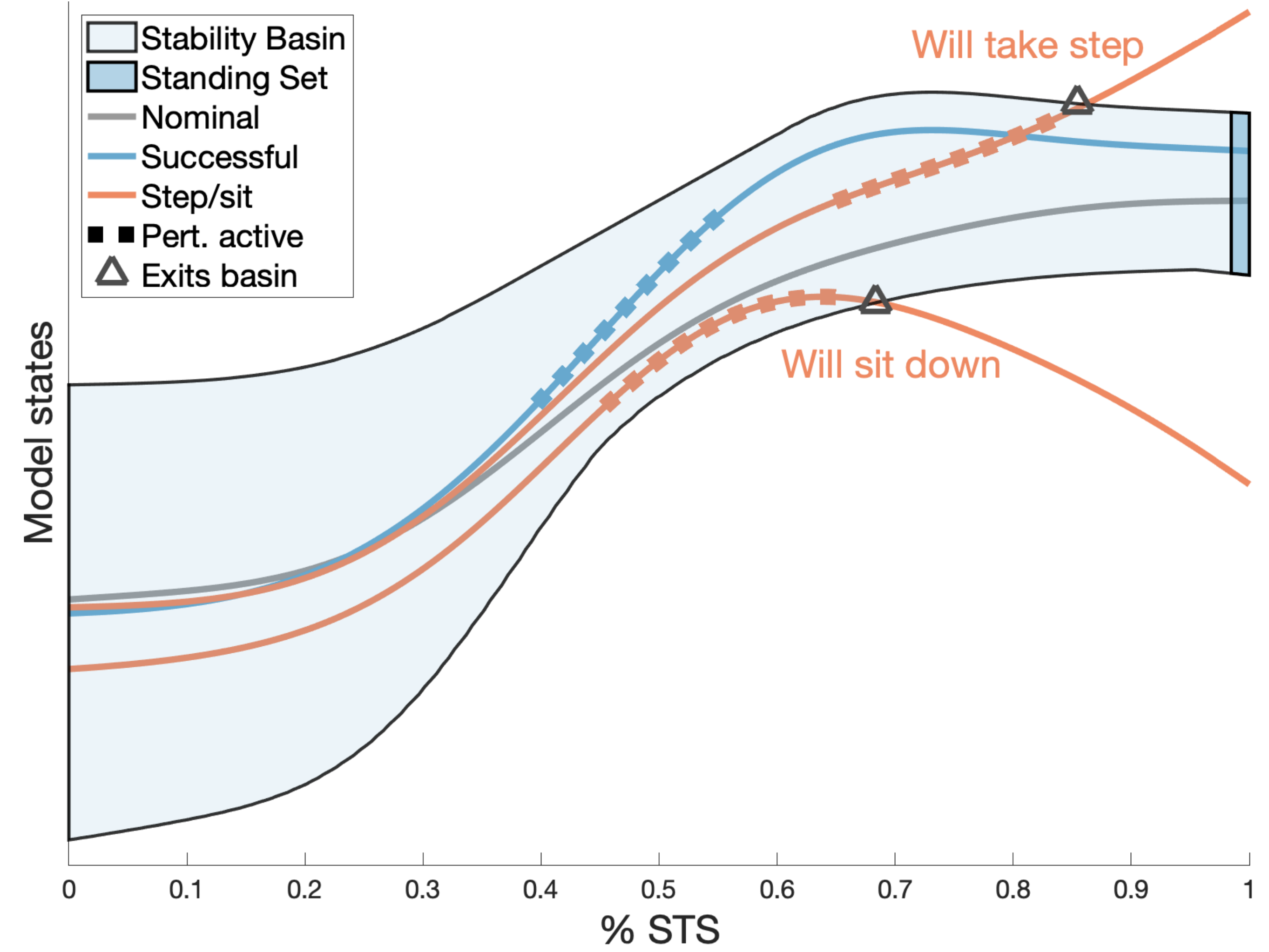}
	\caption{An illustrative overview of SBs for STS.
		The SB represents the set of model states through time that will successfully arrive at a standing set for a given individual and STS strategy.
		Trajectories of the model are illustrated, where the times that perturbations are applied are denoted by the dashed lines.
		Trajectories that exit the SB are predicted to lead to stepping or sitting.}
	\label{fig:intro_fig}
\end{figure}

The goal of this work is to characterize stability during STS.
By stability, we mean the magnitude, direction, and timing of perturbations under which a subject can successfully stand up without stepping, sitting back down, or switching control strategy.
We hypothesize that SBs can accurately represent stability.

The SB can be understood intuitively as follows.
Given a dynamical model of STS, kinematic observations of a subject performing STS can be used to form trajectories of the model.
The SB defines a boundary between ``successful'' and ``unsuccessful'' trajectories of the model, where ``unsuccessful'' implies the subject will have to take a step, sit back down, or switch control strategy.
The SB is introduced pictorially in Fig. \ref{fig:intro_fig}, with a formal definition and computational details given in Sec. \ref{sec:methods:subsec3}.
An accurate SB should predict when a subject's STS motion will be successful or unsuccessful based on model trajectories alone.
To test the accuracy of SBs, we performed a perturbative STS experiment, occasionally inducing subjects to step or sit back down, and then compared SB predictions to experimentally observed outcomes.

This section describes our framework for computing and testing the accuracy of SBs.
First, we collect kinematic observations of a subject performing a perturbative STS experiment (Sec. \ref{sec:methods:subsec1}).
Individualized biomechanical models of STS are constructed for each subject (Sec. \ref{sec:methods:subsec2}).
Strategy-specific input bounds are used to compute the SBs (Sec. \ref{sec:methods:subsec3}).
We test whether the individual and strategy-specific SBs correctly predict when a subject steps or sits down in response to perturbation, and when they are successful (Sec. \ref{sec:methods:subsec4}).
Finally, we perform a comparison to other controller models and a naive method for estimating stability (Sec. \ref{sec:methods:comparisons}).

\subsection{Perturbative STS experiment}
\label{sec:methods:subsec1}

\subsubsection{Experimental protocol}
\label{sec:methods:experimental_protocol}
Subjects began in a seated position on a stool with their arms crossed. 
The height of the stool was adjusted so that the subject's thighs were parallel to the ground.
Subjects practised standing up from the stool and were asked to find a comfortable foot position, which was then demarcated with a line of tape. 

Subjects were instructed to perform three different STS control strategies: their natural strategy, a MT strategy, and a QS strategy as done by Shia et al. \cite{shia2016}.
Subjects watched a demonstration of each strategy, and then practised each strategy a minimum of 10 times prior to data collection.
The following set of treatments were applied to each STS strategy:
\begin{itemize}
\item{Nominal trials: Subjects stood five times from a comfortable foot position using the specified control strategy.}
\item{Foot shift (FS) perturbations: Subjects varied their foot placement in two-inch increments in the anterior-posterior direction from their original position, which were demarcated by taped lines on the ground. 
Subjects stood up at each increment, moving their feet backwards until their heels left the ground, and then forward until a strategy shift was observed.
We noticed that subjects at extreme anterior foot positions attempted to stand up by lunging forward into a squatting position, and excluded these trials from our dataset since they represented a shift in strategy from the subject's nominal behavior.}

\item{Cable pull (CP) perturbations: Subjects returned their feet to their original position. 
A cable system was attached to the subject’s waist and connected to two high-torque motors.
These motor driven cables applied impulses to the subject as they rose.
The cable pulls were restricted to the anterior-posterior direction, and were applied either forward or backwards with variable timing and force.
Specifically, three peak force levels -- low, medium, and high -- were calibrated to each subject.
The low force level was designed to rarely induce stepping or sitting during STS, while the high force level was designed to induce stepping or sitting approximately half of the time.
Six trials were taken at each force level for each STS strategy, with three pulling forwards and three pulling backwards, in random order.
For more details, see Appendix \ref{sec:methods:cable_pull}.}
\end{itemize}

\subsubsection{Kinematic observations} \label{sec:methods:kinematic_observations}
A 10-camera PhaseSpace motion capture system collected kinematic observations of 36 markers at 480 Hz. 
C-Motion's Visual3D biomechanics software was used to fit body segment models to each subject’s data \cite{visual3d}.
MATLAB was used for all subsequent analyses \cite{matlab}.
A 6th-order Butterworth filter with a cut-off frequency of 2 Hz was used to filter joint position trajectories.

We estimate the motion of the subject's center of mass (COM) from these kinematic data.
First, we use a 3-segment model to track the motion of the subject’s shank, thigh, and head-arms-torso segments in the sagittal plane.
The approximate COM of each segment is computed using anthropometric data \cite{winter2005}, and combined to find the trajectory of the total body’s COM position throughout STS.
Then, we obtain COM velocity and acceleration trajectories by numerically differentiating the COM position trajectories.

\subsubsection{Classifying trials as successful/unsuccessful}
\label{sec:methods:classification}
Two clear instances of a subject switching control strategies during STS are taking a step and sitting back down.
We refer to the trials in which these occurred as \defemph{steps} and \defemph{sits}, and collectively label them \defemph{unsuccessful}.
At times, we will also refer to these trials as \defemph{failures}.
We define the onset of stepping or sitting rigorously in Sec. \ref{sec:methods:failureonset}.
If no step or sit is detected, the trial is considered \defemph{successful}.
Note, this scheme can classify trials in which subjects rocked far onto their heels or toes to maintain balance as successful, so long as no step or sit occurs.

\subsection{Dynamic modelling}
\label{sec:methods:subsec2}
To describe our dynamic STS model and trajectory formation, we adopt the following mathematical notation.
Let $A \times B$ denote the Cartesian product of sets A and B.
Let $\mathbb{R}^n$ be the $n$-dimensional Euclidean space, and let $\tau \in [0, T] \subset \R$ denote a time drawn from a time interval.
Without loss of generality, we assume each STS trial starts at time $\tau = 0$.
\subsubsection{Telescoping Inverted Pendulum Model}
\label{sec:methods:modeling}
\begin{figure}[!ht]
	\centering
	\includegraphics[width = 0.5\textwidth]{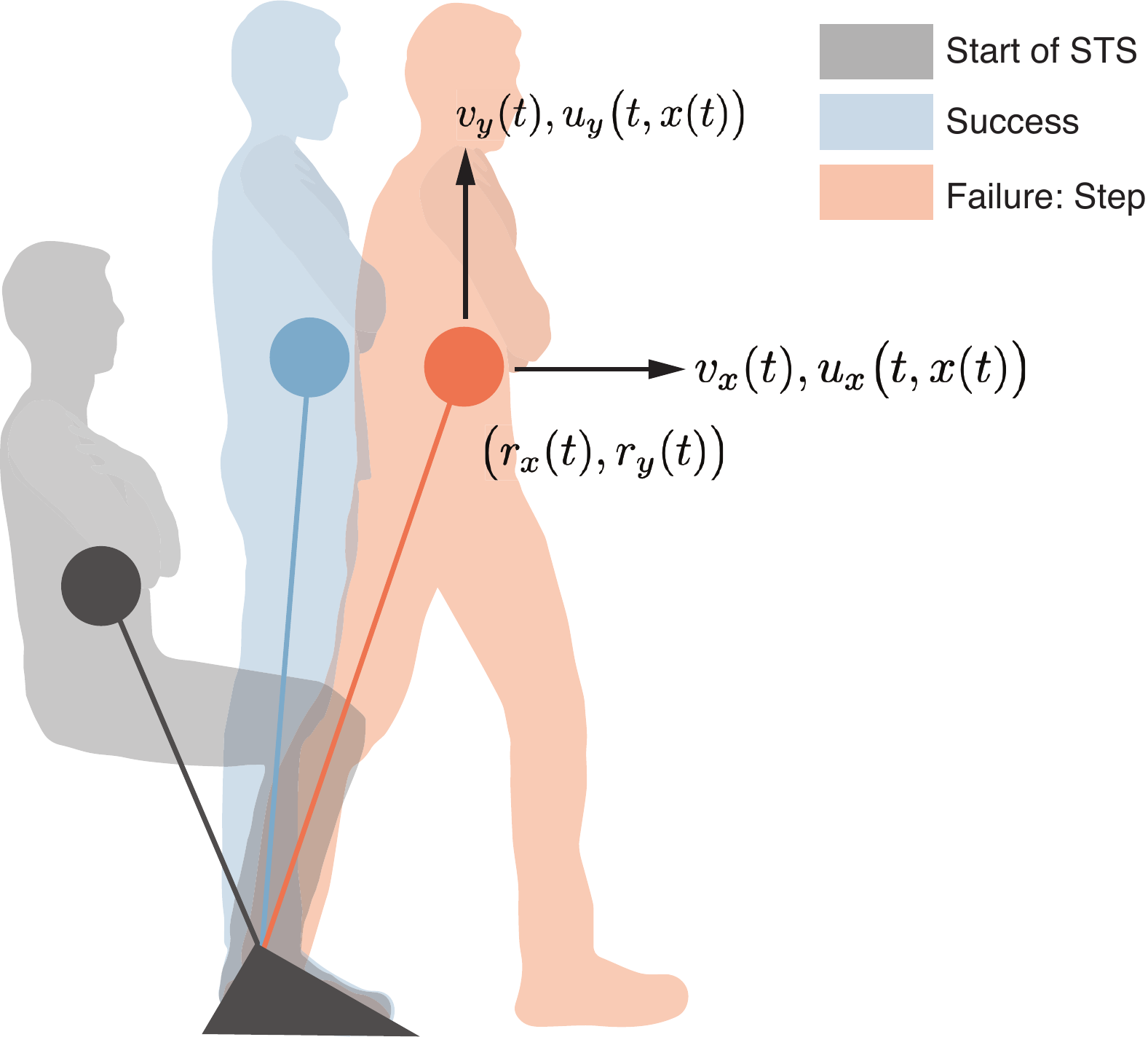}
	\caption{Subjects began from a seated position with their arms crossed against their chests.
	The subject's COM is illustrated by filled circles.
	Cable-pulls applied to the subject sometimes caused them to step or sit; otherwise, the trial was considered to be successful.}
	\label{fig:TIPM}
\end{figure}

We use a Telescoping Inverted Pendulum Model (TIPM, Fig. \ref{fig:TIPM}) to model each subject’s COM motion in the sagittal plane during each STS trial \cite{papa1999}.
This model is capable of describing the sizable displacements of a subject's COM in the horizontal and vertical directions that occur during STS.
The TIPM consists of a point mass of mass $m \in \R$ representing the subject's COM.
Individualized TIPMs are constructed for each subject by setting $m$ as the subject's mass.

We define the origin as the initial position of the ball of the subject's foot.
We let the subscript $(\cdot)_x$ denote a quantity in the anterior-posterior direction, and the subscript $(\cdot)_y$ denote a quantity in the vertical direction.
At each time $\tau \in [0, T]$, we denote the following quantities of the TIPM as:
\begin{itemize}
    \item Positions: $\tilde{r}_x(\tau), \tilde{r}_y(\tau) \in \R$
    \item Velocities: $\tilde{v}_x(\tau), \tilde{v}_y(\tau) \in \R$
    \item Accelerations: $\tilde{a}_x(\tau), \tilde{a}_y(\tau) \in \R$
    \item Inputs: $\tilde{u}_x(\tau), \tilde{u}_y(\tau) \in \R$
    \item Cable-pull forces: $\tilde{d}_x(\tau), \tilde{d}_y(\tau) \in \R$
\end{itemize}

Let $\tilde{x}: [0, T] \to X \subset \R^4$ be a state trajectory, so that $\tilde{x}(\tau) \in X$ is the model's state at time $\tau \in [0, T]$:
\begin{equation}
    \tilde{x}(\tau) = \begin{bmatrix}
    \tilde{r}_x(\tau) \\
    \tilde{v}_x(\tau) \\
    \tilde{r}_y(\tau) \\
    \tilde{v}_y(\tau)
    \end{bmatrix}.
\end{equation}
Let $\tilde{u}: [ 0, T ] \times X \to U \subset \R^2$ be an input trajectory, so that $\tilde{u}(\tau, \tilde{x}(\tau))$ is the input at that time and state:
\begin{equation}
    \tilde{u}(\tau, \tilde{x}(\tau)) = \begin{bmatrix}
    \tilde{u}_x(\tau, \tilde{x}(\tau)) \\
    \tilde{u}_y(\tau, \tilde{x}(\tau))
    \end{bmatrix}.
\end{equation}
The time derivative of $\tilde{x}(\tau)$ at time $\tau$, denoted $\dot{\tilde{x}}(\tau)$, can be written as:
\begin{equation}
\label{eq:TIPM_dynamics}
\dot{\tilde{x}}(\tau) =
\begin{bmatrix}
\tilde{v}_x(\tau) \\
\frac{1}{m} \left(\tilde{u}_x(\tau, \tilde{x}(\tau)) +  \tilde{d}_x(\tau) \right)\\
\tilde{v}_y(\tau) \\
\frac{1}{m} \left(\tilde{u}_y(\tau, \tilde{x}(\tau)) + \tilde{d}_y(\tau) \right) -g
\end{bmatrix},
\end{equation}
where $g$ is the gravitational acceleration $9.81 m/s^2$.

\subsubsection{Nondimensionalizing time} \label{sec:methods:normalizing_trajectories}
The time that it takes to complete STS varies from trial to trial.
Details on how the start and end of each STS trial are chosen are provided in Appendix \ref{sec:methods:segmentation}.
To make accurate comparisons across trials, we normalize each TIPM state trajectory $\tilde{x}$ by the trial's length, so that each trajectory occurs over the interval $[0, 1]$, which we refer to as 0 to 100 \%STS.
We introduce a unitless time variable $t$, related to $\tau$ by $t = \frac{\tau}{T}$.
This dimensionless variable allows us to introduce a normalized state trajectory $x: [0, 1] \to X$, where
\begin{align}
\label{eq:traj_scaling}
    x(t) = \begin{bmatrix}
    \tilde{r}_x(T \cdot t) \\
    T \cdot \tilde{v}_x(T \cdot t) \\
    \tilde{r}_y(T \cdot t) \\
    T \cdot \tilde{v}_y(T \cdot t)
    \end{bmatrix}.
\end{align}
We scale accelerations and forces by a factor of $T^2$ following similar logic.

For the rest of this paper, we only consider normalized trajectories, unless explicitly stated.
Let $r_x, r_y, v_x$ and $v_y$ refer to normalized position and velocity trajectories, and let $a_x, a_y, u_x, u_y$, $d_x$ and $d_y$ refer to normalized acceleration and force trajectories.

\subsection{Computing Stability Basins}
\label{sec:methods:subsec3}
Three elements are required to form the SB for each of an individual's STS control strategies:
\begin{enumerate}
    \item The TIPM dynamics \eqref{eq:TIPM_dynamics}.
    \item Bounds on the TIPM control input denoted $u$, a \defemph{lower bound on the control input} denoted $u^\regtext{lb}$, and an \defemph{upper bound on the control input} denoted $u^\regtext{ub}$.
    Note that
    \begin{equation}
        u^\regtext{lb} \leq u \leq u^\regtext{ub}
        \label{eq:input_bounds_first}
    \end{equation}
    where the inequality is taken element-wise.
    \item A \defemph{standing set}, $X_T$, that encapsulates all model states observed to correspond to successful standing.
\end{enumerate}
Given these elements, we define the SB as follows:
\begin{defn}
\label{def:SB}
The SB $\subset [0, 1] \times X$ is a subset of times (0 - 100\%STS) and model states from which a trajectory of the model obeying the dynamics \eqref{eq:TIPM_dynamics} and input constraints \eqref{eq:input_bounds_first} will arrive at the standing set $X_T$.
\end{defn}

In the rest of this subsection, we explain how we compute the input bounds $u^\regtext{lb}$ and $u^\regtext{ub}$ and the standing set $X_T$ from data.
Then, we use these objects to generate the SBs.

\subsubsection{Bounding control inputs using observations} \label{sec:methods:input_bounds}

\begin{figure*}[!ht]
	\centering
	\includegraphics[width=\textwidth]{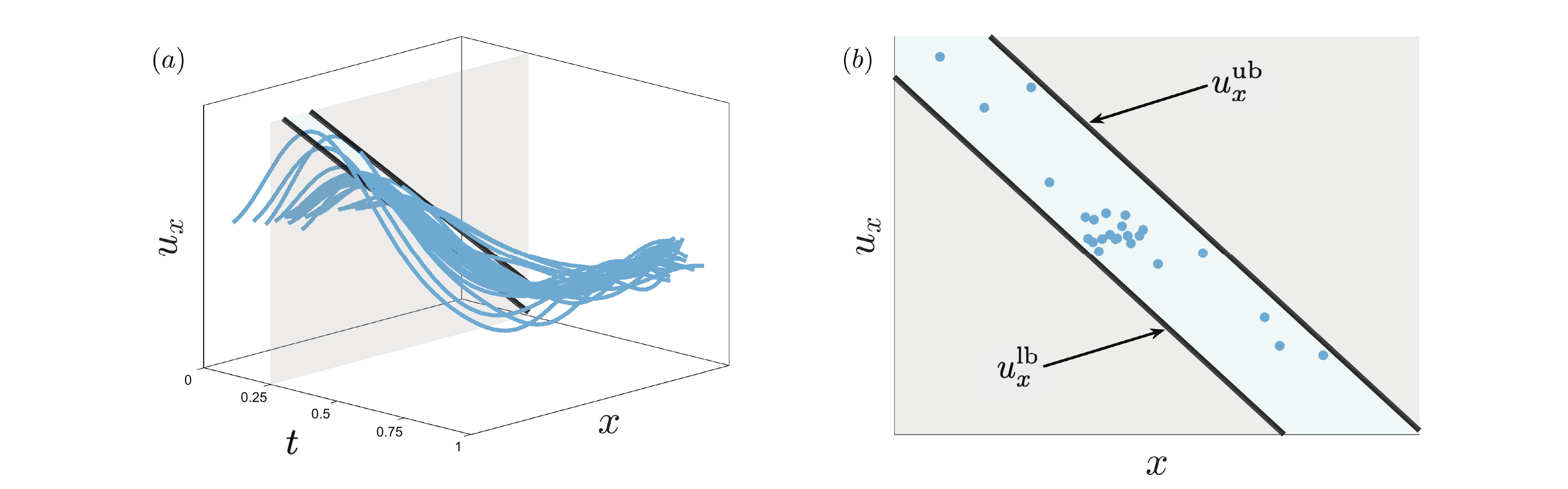}
	\caption{These figures show how we construct the BFF+FB controller from data.
	The process is shown for $u_x$, and is identical for $u_y$.
	In Fig. \ref{fig:input_bounds}(a), $u_x$ evolves as a function of time $t$ and state $x$, which has been depicted as a single dimension for this illustration.
	The blue lines represent the input trajectories of successful trials.
	At a given time $t$, as in Fig. \ref{fig:input_bounds}(b), the upper and lower bounds $u\lb_x(t, x(t))$ and $u\ub_x(t, x(t))$ are parallel linear functions of state that encompass all of the successful inputs.}
	\label{fig:input_bounds}
\end{figure*}

To form the SB, we must specify the TIPM's control input at each time and state. 
Modelling human motion requires this controller to be estimated from observations.
In many previous studies, researchers have modelled human controllers as an open-loop ``feedforward'' component plus a closed-loop ``feedback'' component \cite{kuo2002, todorov2004, shia2016}, finding a single controller that best models the data.
However, since the form of the controller employed by humans is unknown, it is difficult to say with confidence that a single controller adequately captures human behavior.
Therefore, at each instance in time, we compute the range of control inputs of a certain form that can be applied while explaining the data we observed, thereby modelling a range of inputs a human may use to recover from perturbation.
This allows us to compute SBs with a set of hypothetical controllers, instead of just one.
Our method of SB computation, detailed in Sec. \ref{sec:methods:CORA}, is flexible to this type of controller specification, and in particular, we verify that this approach is not too conservative in Sec. \ref{sec:results}.

We present this approach more formally in the following definition:
\begin{defn}
\label{def:BFFFB}
We model the inputs $u$ to the TIPM as a feedforward component plus a linear feedback component, where the feedforward component is drawn from a bounded set.
We refer to this controller model as a bounded feedforward plus feedback \defemph{(BFF+FB)} controller.
Specifically, given time $t$ and state $x(t)$, we define the bounds in \eqref{eq:input_bounds_first} as
\begin{align}
    \label{eq:ulbounds}
    u\lb(t, x(t)) &= \ff\lb(t) - K(t)x(t) \\
    \label{eq:ubbounds}
    u\ub(t, x(t)) &= \ff\ub(t) - K(t)x(t)
\end{align}
where $\ff\lb(t)$ and  $\ff\ub(t) \in \R^2$ are lower and upper bounds on the feedforward input, and $K(t) \in \R^{2 \times 4}$ is a matrix of linear feedback gains.
\end{defn}

We generate the parameters of the BFF+FB controller $\ff\lb, \ff\ub$, and $K$ from data.
This is done separately for each STS strategy.
First, we compute the inputs for each successful trial within that strategy via inverse dynamics using \eqref{eq:TIPM_dynamics} \cite[Chapter 5]{winter1991}. 
Let $x_i(t)$ and $u_{i}(t, x_i(t))$ denote the state and input at time $t$ for the $i^{th}$ successful trial computed via inverse dynamics, and let $S$ denote the set of successful trials within a given strategy.

We solve a constrained linear least squares problem to obtain the bounds $u\lb(t, x(t))$ and $u\ub(t, x(t))$:
\begin{flalign}
\label{eq:input_bounds}
\min_{\ff\lb(t), \ff\ub(t), K(t)} & \phantom{4} \sum_{i\in {S}}\left( u\lb(t, x_i(t)) - u_{i}(t) \right) ^2 + \\
& \hspace*{3cm} + \left( u\ub(t, x_i(t)) - u_{i}(t) \right) ^2 && \nonumber \\
\text{s.t.} & \phantom{4} u\lb(t, x_i(t)) \leq u_{i}(t) && \hspace*{-8ex} \forall i \in  S, \nonumber\\
& \phantom{4} u\ub(t, x_i(t)) \geq u_{i}(t) && \hspace*{-8ex} \forall i \in  S \nonumber
\end{flalign}
where the inequalities are understood element-wise.
This is a quadratic program, which can be solved rapidly and efficiently.
The bounds produced by this program are depicted in Fig. \ref{fig:input_bounds}.

Since many successful trials within an STS strategy involved cable-pull perturbations, we make the following remark:
\begin{rem}
When computing $u_i(t)$ via inverse dynamics for successful CP trials, we discard the portion of the trajectory during which the perturbation was active.
This is possible because the perturbations were time synchronized with the motion capture data.
\end{rem}

\subsubsection{Generating the standing set}
\label{sec:methods:standing}
\begin{figure}[!t]
	\centering
	\includegraphics[width = 0.5\columnwidth]{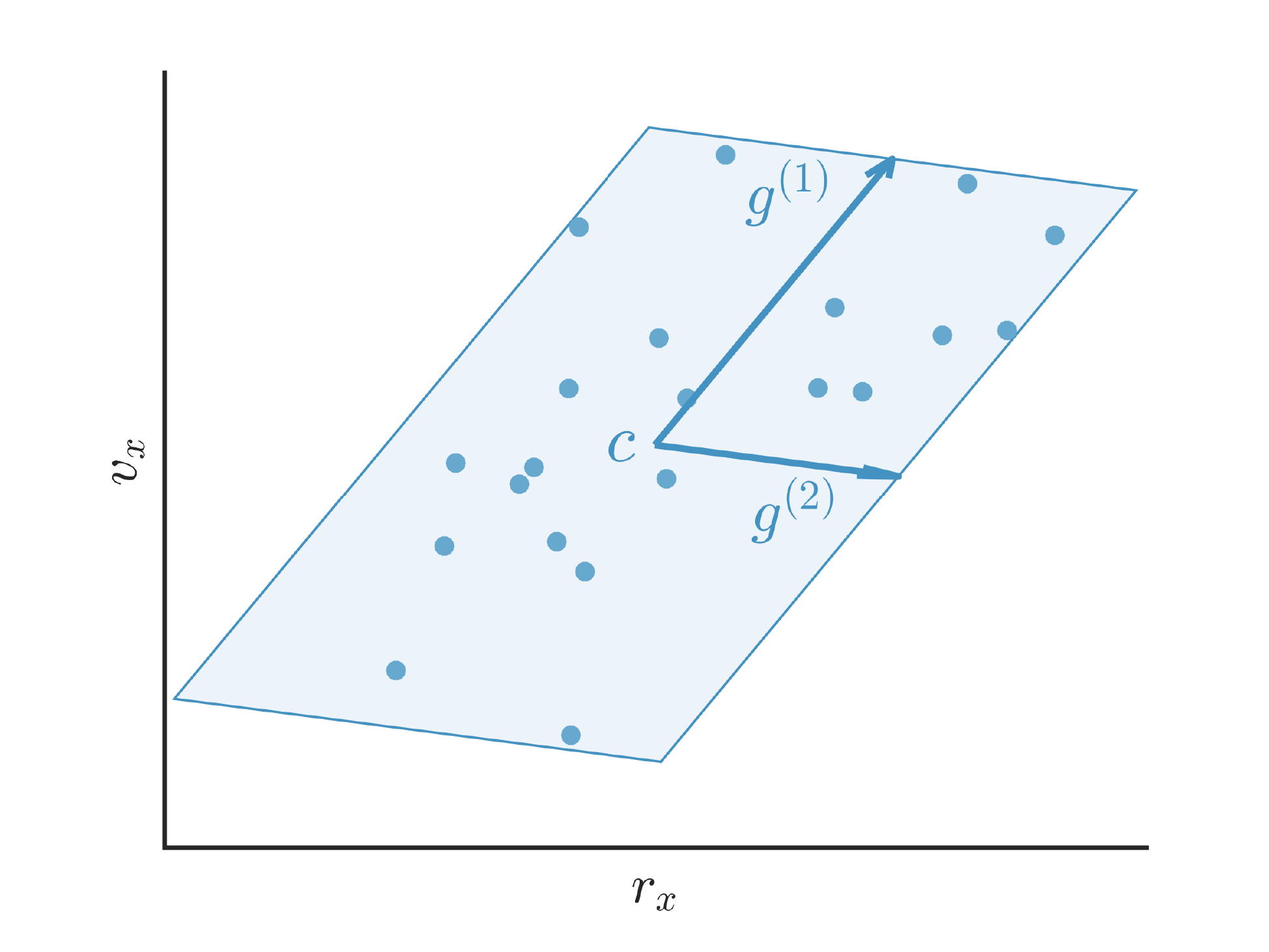}
	\caption{This figure illustrates how $X_T$ is formed from data for a given subject and strategy.
		The states ($r_x$ and $v_x$) at 100\% STS of each observed successful trial within a strategy are illustrated as blue dots.
		An example $X_T$ represented by the shaded blue zonotope contains all of the states observed to correspond to standing.
		The zonotope is parameterized by its center $c$ and generator vectors $g^{(1)}$ and  $g^{(2)}$.
		This example depicts a 2-dimensional $X_T$, but in reality they are 4-dimensional and encompass both horizontal and vertical components of the state space.
	}
	\label{fig:standing}
\end{figure}

The goal of the STS motion is to arrive at a standing configuration, which we represent as the standing set $X_T$.
To define $X_T$, we must first introduce a geometric object called a \defemph{zonotope}.
A zonotope $Z$ is a polytope in $\mathbb{R}^n$ that is closed under linear maps and Minkowski sums \cite{althoff2008}, and is parameterized by its center $c \in \mathbb{R}^n$ and generators ($g^{(1)}, ... g^{(p)}$) shown in Fig. \ref{fig:standing}.
We write the generators as columns of the generator matrix $G \in \mathbb{R}^{n \times p}$.
A zonotope describes the set of points that can be written as the center $c$ plus a linear combination of the generator matrix columns, where each element of the coefficient vector $\beta \in \R^p$ must be between $-1$ and $1$:
\begin{equation}
\label{eq:zono}
Z = \left\lbrace  y \in \mathbb{R}^n \ | \ y = c + G\beta, \quad -\mathbf{1} \leq \beta \leq \mathbf{1} \right\rbrace 
\end{equation}
where the inequalities are applied element-wise and $\mathbf{1}$ is a vector of ones of the appropriate size.

We define $X_T \subset X$ as a zonotope that encompasses the final states (i.e., the states at 100\%STS) of all successful trials observed for a given STS strategy.
Specifically, $X_T$ is a zonotope with 4 generators computed from the final states as proposed by Stursberg \cite[Section~3]{stursberg2003}, where each generator is expanded by 5\% to avoid observed states lying on the edge of the set.
The standing set generation process is depicted in Fig. \ref{fig:standing}.

\subsubsection{Computing SB via reachability analysis}
\label{sec:methods:CORA}

The reader may notice that the SB (Defn. \ref{def:SB}) is defined as a set of states through time that satisfy some dynamics specified by an ordinary differential equation (ODE).
Generally, numerical integration can be used to find solutions of an ODE starting from a single point.
As we are interested in finding all possible solutions of the ODE, subject to constraints, we use a method called reachability analysis to compute the SBs.
Rather than flowing a \textit{single point} through a vector field, reachability analysis flows a \textit{set} through a vector field.
The set of all states that can be attained by the system, starting from some initial set, is referred to as the \textit{forwards reachable set}.
Conversely, all states that can end in some final set are referred to as the \textit{backwards reachable set}.

To compute the SBs, we use an open-source reachability toolbox called CORA, which efficiently handles difficulties associated with representing set dynamics \cite{althoff2015, althoff2016}.
CORA represents the SB as a zonotope at each of a finite collection of \textit{time steps}, which are subintervals of the interval $[0, 1]$ of length $\Delta t$.
With a slight abuse of notation, we let $t$ act as an index, so that $Z\upt$ is the zonotope representing the SB over the time step that contains $t$.
Briefly, CORA works by linearizing the dynamics at each time step about the center of $Z\upt$, and obtaining $Z\uptdt$ by multiplying $Z\upt$ by an overapproximation of the matrix exponential over that time step.
It then expands $Z\uptdt$ to account for the effects of inputs and linearization error.
Each SB was formed using a time step $\Delta t$ of 0.005 (i.e., 0.5\%STS), so that 200 zonotopes represent the SB over the interval $[0, 1]$.

CORA allows the input to the system at each time step to be drawn from a set,
allowing us to use the BFF+FB controller model and let the control input be defined as in \eqref{eq:ulbounds} and \eqref{eq:ubbounds}.
CORA therefore assumes the input at a given time and state can take any value within the input bounds to create the SB.
Since the SB is the set of states through time that arrive at the final standing set $X_T$, it can be computed as a backwards reachable set.
In practice, this means we treat $X_T$ as an initial set, and use CORA to flow the set \textit{backwards} in time under the negative of the dynamics \eqref{eq:TIPM_dynamics}.

\subsection{Evaluating Stability Basin accuracy}
\label{sec:methods:subsec4}
We hypothesize that the SBs can predict when a subject must switch their control strategy in response to perturbation to avoid falling.
Previously in Sec. \ref{sec:methods:classification}, we defined strategy switches as stepping or sitting back down.
In this subsection, we explain how we determine the onset of stepping or sitting.
Then, we use the SBs to predict whether or not a strategy switch will occur by checking if an STS trajectory exits the SB at any point.
Finally, we detail the evaluation procedure we used to test the accuracy of the SBs' predictions.

\subsubsection{Determining onset of failure} \label{sec:methods:failureonset}
To test the predictive power of the SBs, we first identify the onset of failure during an STS movement, $t_f \in [0, 1]$, and show that the SB predicts failure prior to that time.
For step trials, $t_f$ is defined as the time when the subject moves the toes of either foot more than 3 inches in the anterior-posterior direction from its starting point, as measured by motion capture markers placed on the subject’s foot.
Although previous studies measured ground reaction forces to identify the onset of steps \cite{riley1998}, the distance threshold was used here to utilize motion capture data.
For sit failures, $t_f$ is defined as the time when both $v_x$ and $v_y$ become negative, i.e. the COM begins to move back towards the seat. 
If a subject both stepped and sat down, $t_f$ is defined as the earlier of the two.

Since we are using a procedure based on an average nominal trial for aligning and segmenting trials (detailed in Appendix \ref{sec:methods:segmentation}), it is possible for the onset of stepping or sitting to occur after the trial end time $t = 1$.
The end times chosen by the segmentation procedure do not necessarily imply that the subject has reached standing, but are just a means to achieve a consistent segmentation of the data.
We consider a trial with a step or sit occurring later than the trial's end to be unsuccessful, and set $t_f = 1$ in this case.

\subsubsection{Checking trajectories} \label{sec:methods:checking_trajectories}
\begin{figure}[t]
	\centering
	\includegraphics[width = 0.5\columnwidth]{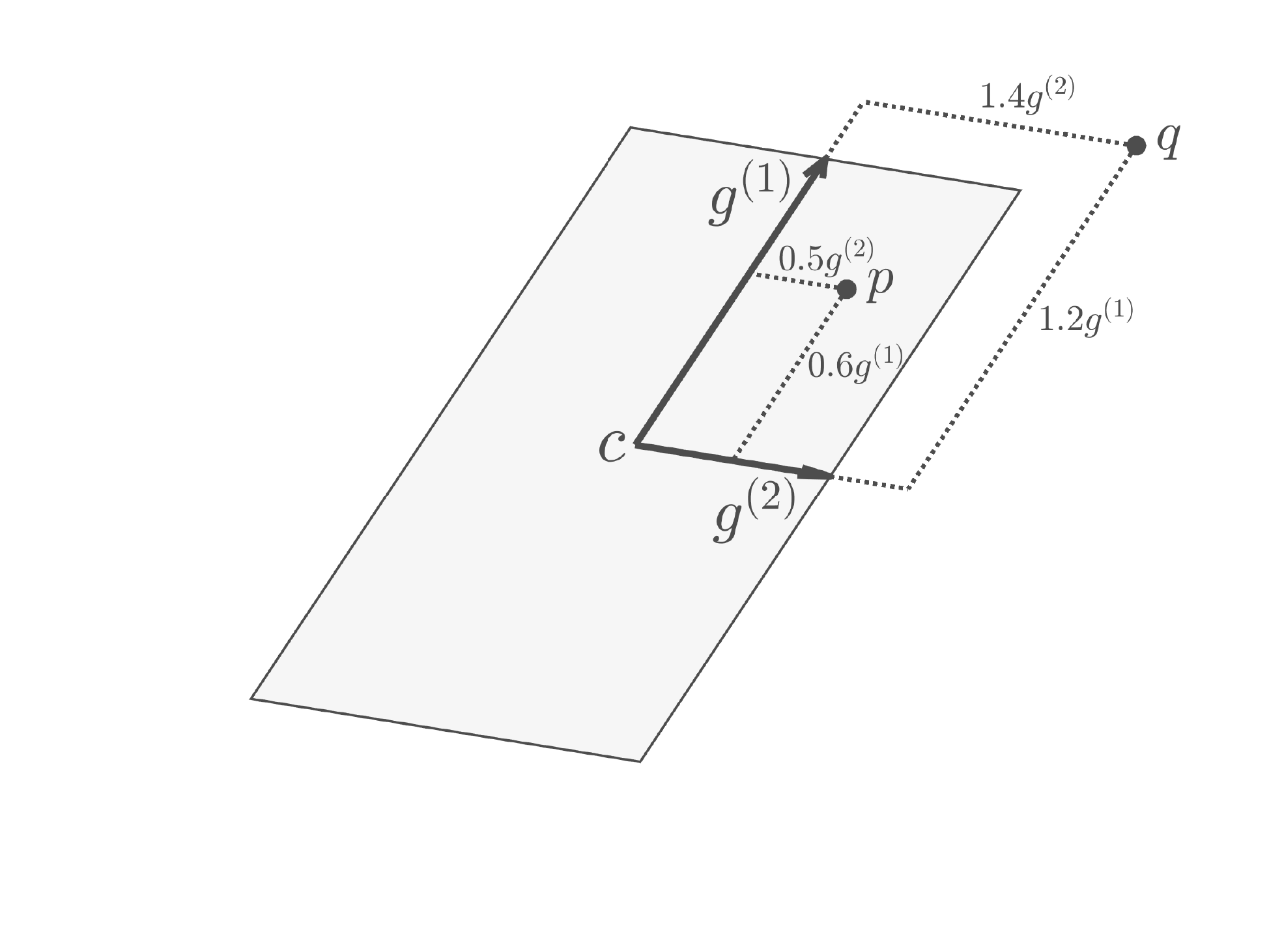}
	\caption{This figure illustrates how to check if a point is inside or outside of a zonotope.
	An example zonotope (shaded grey) is parameterized by its center $c$ and generator vectors $g^{(1)}$ and  $g^{(2)}$.
	A test point $p$ is contained within the zonotope, since the maximum absolute value of the coefficients on $g^{(1)}$ and $g^{(2)}$ is less than or equal to $1$.
	Another test point $q$ lies outside of the zonotope, since the maximum absolute value of the coefficients on $g^{(1)}$ and $g^{(2)}$ is greater than $1$}
	\label{fig:inout}
\end{figure}
Given a trajectory $x$ from an STS trial, we use a linear program to determine whether that trajectory lies inside the SB at each time step.
Recall that each zonotope $Z\upt$ representing the SB is parameterized by its center $c\upt \in \R^4$ and generator matrix $G\upt \in \R^{4 \times p}$. 
At each time $t$, we represent the state $x(t)$ as a linear combination of the generator matrix's columns, and use a linear program to find the representation that has the smallest maximum coefficient $\beta\upt \in \R$:
\begin{flalign}
\label{eq:LP}
\min_{\beta\upt \in \R, \beta \in \R^p} & \phantom{4} \beta\upt && \\
\text{s.t.} & \phantom{4} G\upt \beta = x(t) - c\upt, &&  \nonumber\\
& \phantom{4} |\beta| \leq \beta\upt && \nonumber
\end{flalign}
where the absolute value and inequalities are applied element-wise.
If $\beta\upt$ is less than or equal to 1, the point $x(t)$ is inside the SB at that time step, depicted in Fig. \ref{fig:inout}.

\subsubsection{Evaluation procedure} \label{sec:methods:validation}
The SBs are used to predict the success or failure of every trial within each STS strategy.
The accuracy of each SB is then tested by comparing its predictions to the experimentally observed outcomes of each STS trial.

Given the state trajectory $x$ observed for an STS trial, the linear programs described in \eqref{eq:LP} are used to predict success/failure.
If a state trajectory $x$ remains inside the SB at each time step, the trial is predicted to be successful.
If $x$ exits the SB at any time step, the trial is predicted to fail.
To ensure that the SBs' predictions are made fairly, we make the following restrictions:
\begin{enumerate}
    \item For CP trials, the prediction uses only the portion of the SB after the onset of perturbation.
    \item For trials in which failure was observed, the prediction uses only the portion of the SB before failure occurred (i.e., $t \leq t_f$ as in Sec. \ref{sec:methods:failureonset}).
\end{enumerate}

We perform separate types of evaluation for successful trials versus unsuccessful trials.
For successful trials, we employ a \defemph{leave-one-out} procedure to avoid training and testing on the same data.
After forming the standing set, we leave one successful trial out of the BFF+FB computations in \eqref{eq:input_bounds}.
We then construct an SB using this BFF+FB controller to test the trial that was left out.
We repeat this procedure until all successful trials have been tested.

For unsuccessful trials, we form a single SB using a BFF+FB controller formed using all observed successful trials.
We use this single SB to predict the outcome of each unsuccessful trial within a given strategy.

\subsection{Comparisons} \label{sec:methods:comparisons}
We compare SBs formed using the BFF+FB controller model (Defn. \ref{def:BFFFB}) to two other controller models, as well as a naive method for estimating stability.
The same validation procedure described in Sec. \ref{sec:methods:validation} is performed for each.

\subsubsection{LQR controller}
Previously, Shia et al. proposed to model STS using an LQR controller about a nominal STS trajectory \cite{shia2016}.
To build this controller, we first form a single average nominal trajectory $\bar{x}$ for each of a subject's STS control strategies by taking the mean of the five nominal STS trajectories.
We generate an open loop controller, $u_{ol}: [0, 1] \to U \subset \mathbb{R}^2$, for each average nominal trajectory via inverse dynamics.
Then, we use LQR to design a linear feedback controller about the average nominal trajectory.
We use a quadratic cost function specified by the state weighting matrix $Q = I_{4x4}$ and input weighting matrix $R = 1x10^{-4} I_{2x2}$ to generate the feedback matrix $K$ \cite[Chapter 16]{lancaster1995}, where the weighting matrices were found empirically to produce the best results.
At a given time and state, the input from the LQR controller can then be written as:
\begin{equation}
    u(t, x(t)) = u_{ol}(t) - K(x(t) - \bar{x}(t))
\end{equation}

\subsubsection{FF + FB controller}
We also test the efficacy of a traditional feedforward plus feedback controller (FF+FB), mentioned in \ref{sec:methods:input_bounds}.
This controller differs from the BFF+FB controller because the feedforward component is not drawn from a set.
Let the input $u(t, x(t))$ at some time $t$ and state $x(t)$ be described by the state feedback matrix $K(t) \in \mathbb{R}^{1 \times 4}$ and a feedforward component $\ff(t)$:
\begin{equation}
    u(t, x(t)) = \ff(t) - K(t)x(t)
\end{equation}
To generate $K(t)$ and $\ff(t)$ from data, we use linear least squares to solve a program very similar to \eqref{eq:input_bounds}:
\begin{flalign}
\min_{\ff(t), K(t)} & \phantom{4} \sum_{i\in {S}}\left( u(t, x_i(t)) - u_{i}(t) \right) ^2 && \label{eq:FF+FB}
\end{flalign}
As in \eqref{eq:input_bounds}, $S$ represents the set of observed successful trials within a STS strategy, and $x_i(t)$ and $u_{i}(t)$ correspond to the observed states and input of the $i^{\text{th}}$ trial at time $t$.

\subsubsection{Naive method}
Stability can be estimated from observed perturbed trials by simply drawing a volume around the state trajectories of the observed successful trials.
This method does not utilize reachability or a controller model, and relies on state trajectories alone.
Here, we apply this method by generating a 4-dimensional zonotope at each time step that encloses all of the observed successful states.
This is the same procedure that we employ to generate the standing set $X_T$ (described in Sec. \ref{sec:methods:CORA} and illustrated in Fig. \ref{fig:standing}) applied at each instance in time over the course of the motion.
Since this estimate of stability is constructed using zonotopes, we follow the same leave-one-out validation procedure described in Sec. \ref{sec:methods:validation}.

\section{Results} \label{sec:results}

\begin{figure}[!t]
	\centering
	\includegraphics[width = 0.5\columnwidth]{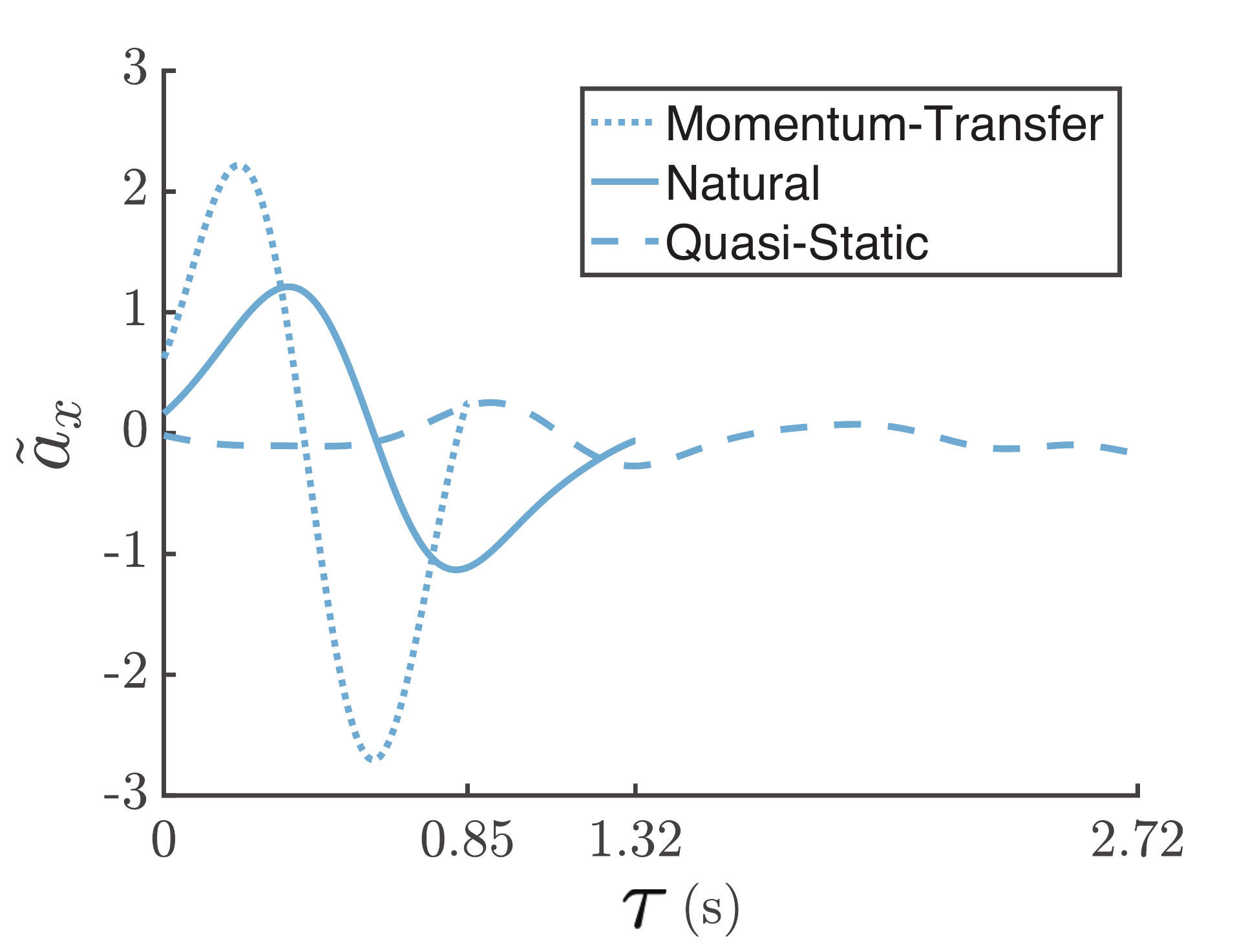}
	\caption{This figure shows example nominal trajectories for each STS strategy for subject ID 8.}
	\label{fig:strategy_ax}
\end{figure}

\begin{figure*}[!htb]
	\centering
	\includegraphics[width=\textwidth]{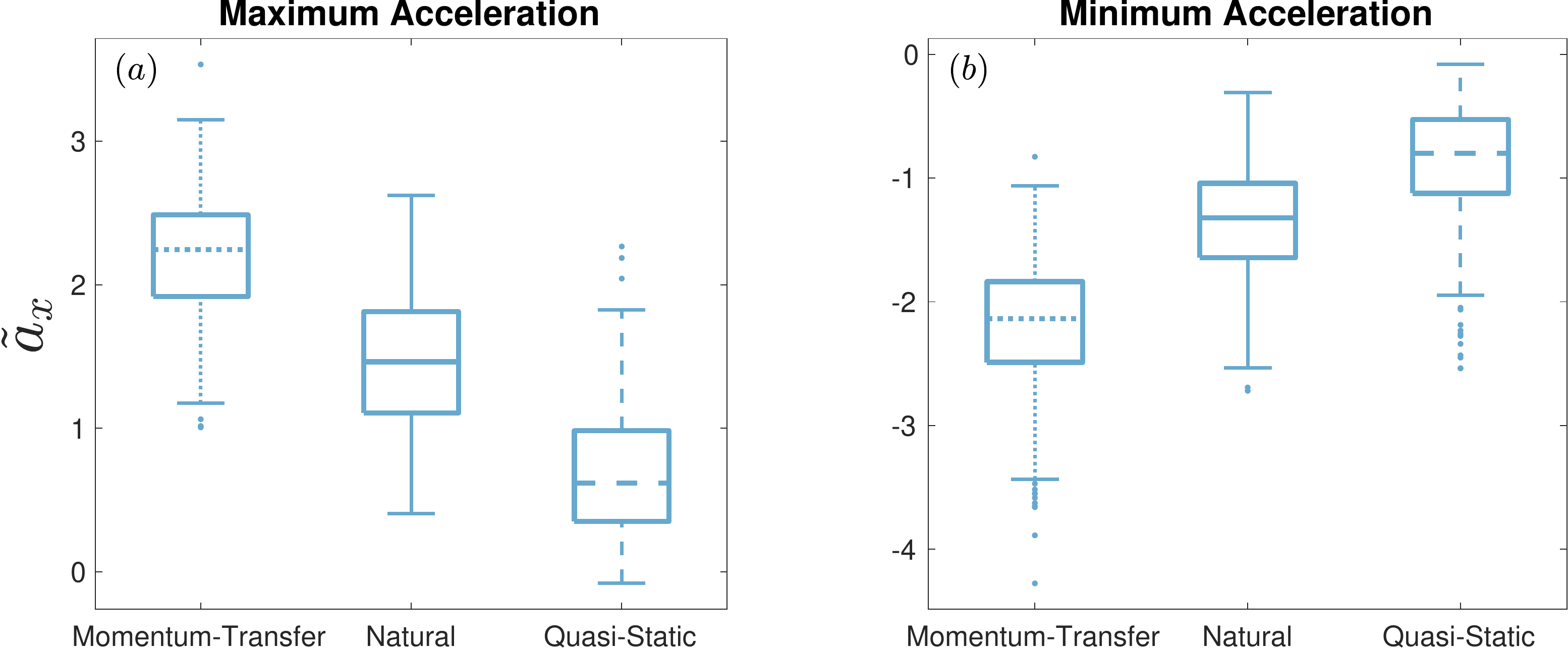}
	\caption{Maximum (Fig. \ref{fig:strategy_stats}(a)) and minimum (Fig. \ref{fig:strategy_stats}(b)) horizontal accelerations are significantly different between STS strategies, with $p < 0.001$ when comparing the maximums or minimums of any two strategies under a two-sample t-test.}
	\label{fig:strategy_stats}
\end{figure*}

We collected 948 trials from 11 participants (three female and eight male; ages 18-32; height 1.70 $\pm$ 0.12 m; body mass 65.4 $\pm$ 10.2 kg), including 213 unperturbed, 194 foot-shifted, and 591 cable pull trials.
Each subject gave their informed written consent, and had no physical or balance disorders which could affect their ability to perform STS. 
The subjects performed STS trajectories with generally the same characteristics as the (different) subjects examined by Shia \cite{shia2016}.
Even though subjects were allowed to determine the speed with which each STS was performed, each subject's natural, MT, and QS STS strategies showed distinct kinematic characteristics, depicted in Fig. \ref{fig:strategy_ax}.
In particular, the maximum and minimum horizontal accelerations of the COM for the MT ([-2.1, 2.2] m/s$^2$) and QS ([-0.8, 0.6] m/s$^2$) strategies were significantly different from each other as well as from the natural strategy ([-1.3, 1.5] m/s$^2$), depicted in Fig. \ref{fig:strategy_stats}, with p-values $< 0.001$ when comparing any two strategies.
While the duration of natural and MT trials were similar (average times of 1.25s and 1.13s), QS trials lasted longer (average 2.23s), due in part to the low observed COM accelerations.
The SB shape and volume are highly dependent on the kinematic and temporal characteristics corresponding to the strategy selected, as in previous work \cite{shia2016}.

All foot-shift trials were considered successful.
When using the MT strategy, subjects were able to stand with their feet further forward (max. $7.82 \pm 2.75$ inches) than when using QS (max. $1.64 \pm 1.21$ inches).
Subjects attempting the extreme anterior and posterior foot positions were either unable to initiate the STS movement, or violated the parameters of acceptable trials stated in Sec. \ref{sec:methods:experimental_protocol}.

Cable pull perturbations during STS frequently induced a step or sit to avoid falling.
Out of 591 cable pull trials, we observed failure in 198, or 33.5\%.
Specifically, we observed 42 steps and 29 sits for the natural strategy, 40 steps and 19 sits for MT, and 39 steps and 29 sits for QS.
Fewer sits were observed for the MT strategy than in both the natural and QS strategies, likely due to the larger forward momentum at seatoff for the MT strategy.

\begin{table*}[!htb]
\centering
\caption{Statistics collected from the perturbative STS experiment are summarized below.
Note that the onset of failure (Step \%STS and Sit \%STS) often occurred after the defined trial end, which is discussed in Sec. \ref{sec:methods:failureonset}.
} \label{tab:expstats}
\begin{tabular}{c | c c c c c c c c}
 STS Strategy & Total & CP & Steps & Sits & Trial Time (s) & Pert. Onset \%STS & Step \%STS & Sit \%STS \\
 \hline
 Natural & $326$ & $197$ & $42$ & $29$ & $1.25 \pm 0.23$ & $50.34 \pm 9.94$ & $107.27 \pm 20.92$ & $96.16 \pm 15.35$\\
 MT & $322$ & $198$ & $40$ & $19$ & $1.13 \pm 0.21$ & $46.58 \pm 11.03$ & $114.30 \pm 32.66$ & $99.38 \pm 15.21$\\
 QS & $300$ & $196$ & $39$ & $29$ & $2.23 \pm 0.89$  & $50.87 \pm 16.71$ & $95.04 \pm 20.53$ & $85.26 \pm 20.14$\\
\end{tabular}
\end{table*}

\begin{figure*}[!p]
    \centering
    \includegraphics[width = 0.86\textwidth]{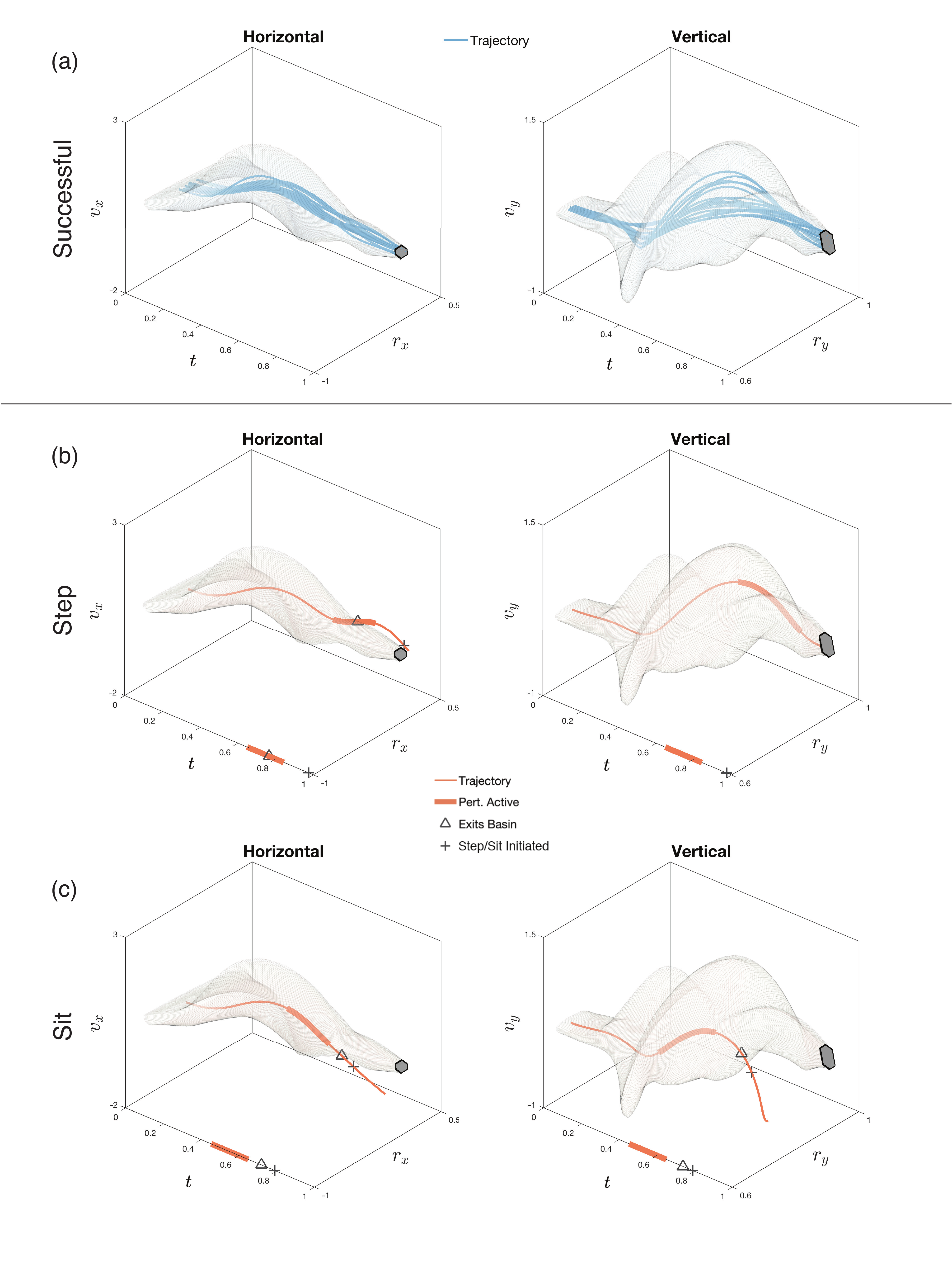}
    \caption{The computed natural strategy SB for subject ID 6.
    The horizontal and vertical projections of the SB are represented as the regions encapsulated by the light grey borders.
    The projection of the standing set $X_T$ is shown as the dark grey region on the right side of each plot.
    The state trajectories of all of subject ID 6's successful natural strategy trials remain inside the SB at all times, as shown in Fig. \ref{fig:all_basins}(a).
    In contrast, state trajectories of a step (Fig. \ref{fig:all_basins}(b)) and a sit (Fig. \ref{fig:all_basins}(c)) exit the basin before the onset of failure.
    As detailed in Sec. \ref{sec:methods:classification}, we define step initiation as the time when the toes of either foot move more than 3 inches in the anterior-posterior direction, and sit initiation as the time at which both $v_x$ and $v_y$ become negative.
    }
\label{fig:all_basins}
\end{figure*}

SBs were generated on a laptop computer with a 2.7 GHz Intel Core i7 processor.
SBs computed using the BFF+FB controller take $0.76 \pm 0.014$ seconds to compute.
To assess the accuracy of each SB, we compared the predictions to the experimental observation for each trial according to the procedure described in Sec. \ref{sec:methods:validation}.
The aggregate prediction rates of the SBs formed using the BFF+FB controller for each STS strategy across subjects are given in Tab. \ref{tab:sbresults}.
Across STS strategies, successful trials (Fig. \ref{fig:all_basins}(a)) remain inside the basin 95.73\% of the time.
Of the failure trials, 90.91\% of steps are predicted prior to onset (Fig. \ref{fig:all_basins}(b)), while 92.21\% of sits are predicted prior to onset (Fig. \ref{fig:all_basins}(c)).
There was no consistent effect of perturbation force on the predictive power of the method, detailed in Tab. \ref{tab:sbCPresults}.
\begin{table*}[!htb]
\centering
\caption{The accuracy of the SBs generated using the BFF+FB controller model are reported below.
This table reports a tally of the correct predictions for each STS strategy and trial type across subjects.
} \label{tab:sbresults}
\begin{tabular}{ c | c c c }
 STS Strategy & Successful Trials & Step Trials & Sit Trials \\
 \hline
 Natural & 248/255 -- 97.25\%  & 41/42 -- 97.62\% & 27/29 -- 93.10\% \\
 MT & 257/263 -- 97.72\% & 33/40 -- 82.50\% & 17/19 -- 84.75\% \\
 QS & 213/232 -- 91.81\% & 36/39 -- 92.31\% & 27/29 -- 93.10\% \\
 \hline
 Combined & 718/750 -- 95.73\% & 110/121 -- 90.91\% & 71/77 -- 92.21\% \\ 
\end{tabular}
\end{table*}

\begin{table*}[!htb]
\centering
\caption{The accuracy of SBs generated using the BFF+FB controller model are reported for CP trials at three different force levels of the applied perturbation. The results are aggregated across subjects and the three tested STS control strategies.}
\label{tab:sbCPresults}
\begin{tabular}{ c | c c c }
 CP Force Level & Successful Trials & Step Trials & Sit Trials \\
 \hline
 Low & 170/176 -- 96.59\%  & 15/15 -- 100\% & 4/5 -- 80.00\% \\
 Medium & 125/132 -- 94.70\% & 34/38 -- 89.47\% & 24/27 -- 88.89\% \\
 High & 80/85 -- 94.12\% & 61/68 -- 89.71\% & 43/45 -- 95.56\% \\
\end{tabular}
\end{table*}

Overall, the SB failure predictions are over 90\% accurate; only 17 out of 198 unsuccessful trials are incorrectly predicted to be successful.
To understand these false successes, we measured the maximum Euclidean distance of each unsuccessful trajectory to its nearest successful trajectory.
The false successes are 34.24\% closer to their nearest successful trajectories than are the true failure trajectories.
Additionally, the onset of failure with respect to the onset of perturbation for false successes averaged 9.47\% earlier than in the true failures. 
Note that it is plausible that individuals switched control strategies for some subset of the trials (e.g., from QS to natural), which could affect the accuracy of the strategy-specific SBs.
Since we only characterize trials that are sits or steps as failures, it is possible that an individual switched strategies to successfully stand but would not have reached standing with their original strategy.

\begin{figure*}[t]
    \centering
    \includegraphics[width=\textwidth]{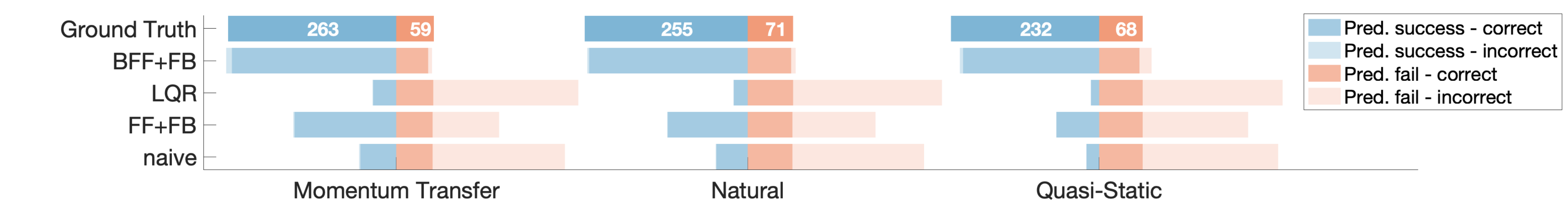}
    \caption{This figure compares the predictive accuracy of the SBs formed using the BFF+FB controller to three other methods.
    Results are presented for each STS strategy, aggregated across subjects.
    The BFF+FB controller's predictions for both successes and failures are near the ground truth, while the three other methods predict many more failures than were experimentally observed.}
    \label{fig:results}
\end{figure*}

We compared the accuracy of SBs computed using the BFF+FB controller to SBs formed using other methods (Tab. \ref{tab:compresults}.
SBs formed using the LQR controller proposed in previous work \cite{shia2016} correctly predict 9.47\% of successful trials.
SBs formed using a traditional FF+FB controller correctly predict the outcome of 46.80\% of successful trials.
Finally, a naive method for estimating stability yields an accuracy of 16.67\% for successful trials.
Though each of the LQR, FF+FB, and naive methods yield nearly 100\% prediction accuracy for trials where failure occurred, the low successful trial prediction rates mean that few trials remain within each SB, implying a severe underapproximation of the true stable region.
We also give the false successful prediction and false failure prediction rates of each method in Tab. \ref{tab:compresults}, and display these rates for each strategy in Fig. \ref{fig:results}.
Each of the LQR, FF+FB, and naive approaches for computing SBs has a high false failure prediction rate, further showing that each underestimates the size of the true stable region.

\begin{table*}[t]
\centering
\caption{The accuracy of the SBs formed using the BFF+FB controller model were compared to SBs formed using other methods. The results of the comparison are reported below, and are aggregated across subjects and STS strategies.}
\label{tab:compresults}
\begin{tabular}{ c | c c c c }
 Basin Type & Successful Trials & Step/Sit Trials & False Successful Predictions & False Step/Sit Predictions \\
 \hline
 BFF+FB & 718/750 -- 95.73\%  & 181/198 -- 91.41\% & 17/735 -- 2.31\%  & 32/213 -- 15.02\% \\
 LQR & 71/750 -- 9.47\% & 197/198 -- 99.49\% & 1/72 -- 1.39\%  & 679/876 -- 77.51\% \\
 FF+FB & 351/750 -- 46.80\% & 195/198 -- 98.48\% & 3/354 -- 0.85\%  & 399/594 -- 67.17\% \\
 naive  & 125/750 -- 16.67\% & 195/198 -- 98.48\% & 3/128 -- 2.34\%  & 625/820 -- 76.21\% \\
\end{tabular}
\end{table*}

\section{Discussion} \label{discussion}
In theory, stability can be directly characterized by perturbing a subject in an infinite number of ways and drawing a volume around the observed successful trials.
Because this approach is practically infeasible, several model-based metrics have been developed for assessing dynamic stability.
Many, such as Floquet Multipliers and Lyapunov Exponents, are limited to analyzing small perturbations during periodic motion.
Only a few metrics assess the largest perturbations that can be withstood during motion \cite{bruijn2014}, or the stability of aperiodic motion.
In this area, Fujimoto and Chou \cite{fujimoto2014} constructed estimates of the stable region for STS at a single time based on the extrapolated center of mass concept.
We build on these methods by leveraging novel data-driven controller models, reachability analysis, and subject-specific dynamics to estimate stability over the course of the STS motion.
By performing a perturbative experiment, we demonstrate that our SB-based approach accurately predicts stability across a diverse set of body morphologies, perturbations, and STS control strategies.

Our data-driven control model consists of a feedforward (``predictive’’) and a linear feedback (``reactive’’) component.
Although there is evidence that humans employ feedforward and feedback terms for control \cite{kuo2002}, the exact form of these controllers remains unclear.
Linear feedback applied about the COM has been sufficient to explain observed walking \cite{wang2014} and standing \cite{park2004} behavior.
However, we found the LQR controller proposed by Shia et al. \cite{shia2016} mischaracterizes the size and shape of the stable region because it is not trained on any of the observed perturbed trials.
The classical FF+FB controller produced false failures, implying that this underestimates the true stable region of movement.
Instead, our BFF+FB controller model draws from a set of feedforward inputs, plus a linear feedback term, that encapsulates all inputs from observed successful trials, resulting in improved performance over models that return a single input.
This improvement in the accuracy of stability estimation is an important step towards clinical deployment.

In addition to increased accuracy, the computationally-tractable, individualized, and time-varying aspects of the SB approach make it well suited for widespread deployment and integration with existing clinical methods.
The rapid and automated computation of SBs from kinematics alone minimizes the temporal and financial investment in performing this analysis, especially if coupled with computer vision techniques for pose estimation from digital cameras \cite{cao2017}.
This can enable longitudinal studies examining how stability changes over time, and whether a smaller SB is correlated with fall risk, to be performed both accurately and easily.
Although we make use of perturbative data for computing SBs here, one advantage of our proposed method is that it only relies on data from successful trials, and does not require a subject to be perturbed to failure.
Thus, the SB method shows promise for characterizing the stability of subjects already at a high risk of falling or injury, such as elderly or rehabilitating subjects.
Additionally, the TIPM and BFF+FB models are computed for each individual, enabling examination of a wide range of body morphologies and STS control strategies.
Furthermore, stability throughout a task can be studied by measuring the size and location of cross-sections of the SB at particular times.
By combining the SB approach with other clinical techniques, such as electromyography, clinicians can identify muscles associated with unstable portions of motion to target preventative and rehabilitative care.
Such detailed analyses will undoubtedly provide deeper insight into the mechanisms underlying motor control.

Our approach can also be used in real time to inform controller design for wearable robotic devices because it provides time-varying boundaries on the region inside which an individual is stable.
If a prosthetic device or exoskeleton detects that its user has exited the SB and become unstable, it can adapt its control scheme to aid in recovery.
Furthermore, just as current tuning of controllers for prosthetic devices and exoskeletons focuses on minimizing metabolic cost, \cite{brockway1987, koller2016}, wearable robotic controllers can be optimized to increase stability by maximizing the volume of the SBs for the combined robot-human system. 
Thus, the high time-varying accuracy and low investment in computation demonstrate that the SB approach has the potential to increase access of personalized preventative and rehabilitative mobility care and inform the design of robotic systems for stable movement.

\FloatBarrier
\section*{Ethics}
The experimental protocol was approved by the University of Michigan Health Sciences and Behavioral Sciences Institutional Review Board, eResearch ID: HUM00020554.

\section*{Data, code and materials}
Code and STS data may be found here: \href{https://github.com/pdholmes/STS_SB}{https://github.com/pdholmes/STS\_SB}

\section*{Competing interests}
We declare we have no competing interests.

\section*{Authors' contributions}
PDH conducted the experiment, analysed the data, wrote SB computation and validation code, and drafted the manuscript;
SMD analysed the data, wrote SB computation code, and drafted the manuscript;
XYF constructed perturbation equipment, conducted the experiment, and edited the manuscript;
TYM conducted the experiment, supervised data analysis, and edited the manuscript;
RV conceived and directed the study, and edited the manuscript.

\section*{Acknowledgements}
We thank Art Kuo for helping to collect this dataset, and Victor Shia for his code and advice.
We thank the 11 anonymous subjects for their participation in the study.

\section*{Funding}
This work is supported by the National Science Foundation, under CAREER Award 1751093 and Graduate Research Fellowship Program Grant No. 1256260 DGE.

\bibliographystyle{unsrt}
\bibliography{references}

\clearpage

%%TC:ignore
 \appendix
 \section{Generating cable pull perturbations} \label{sec:methods:cable_pull}
Two high-torque motors were attached to cable-pulley systems and placed in front of and behind the test platform. 
The cables were attached to a belt around the subject’s waist, and the height of the pulleys were adjusted such that the cables were horizontal when the subject was standing.
The pulleys were located approximately 3-4 feet in front of and behind the subjects.
The cables were only attached for the CP condition, and detached for the nominal and FS conditions.

A custom-written LabVIEW \cite{labview} program was used to control the motor torques. 
During the experiment, a low torque was constantly commanded in both motors to keep the cables from going slack.
The torque level was balanced between the anterior and posterior cables to minimize force bias felt by the subject.
Force sensors on the cables were synchronized with motion capture data, so that the timing of the perturbation application in relation to kinematic data was known. 
Perturbations were manually activated by the toggling of a handheld switch.
The experiment operator applied perturbations to the subject at variable times following seatoff to gather a range of recovery responses over the course of STS.

Perturbations were active over a 250 ms period, during which the applied force would ramp up to a specified peak, and then quickly ramp down to a baseline. 
The peak force to be applied was determined before the experiment began.
Specifically, three peak force levels -- low, medium, and high -- were calibrated to each subject.
The low force level was designed to rarely induce stepping or sitting during STS, while the high force level was designed to induce stepping or sitting approximately half of the time.
Stepping and sitting are more formally defined in Sec. \ref{sec:methods:failureonset}, and we report the number of steps and sits observed for each force level in Sec. \ref{sec:results}.
These force levels were roughly chosen based on the subject’s height and weight, and adjusted manually during a pre-experiment test session. 
Six trials were taken at each force level for each STS strategy, with three pulling forwards and three pulling backwards, in random order.

\section{Determining start and end of STS} \label{sec:methods:segmentation}
\begin{figure}[htb]
	\centering
	\includegraphics[width = 0.5\columnwidth]{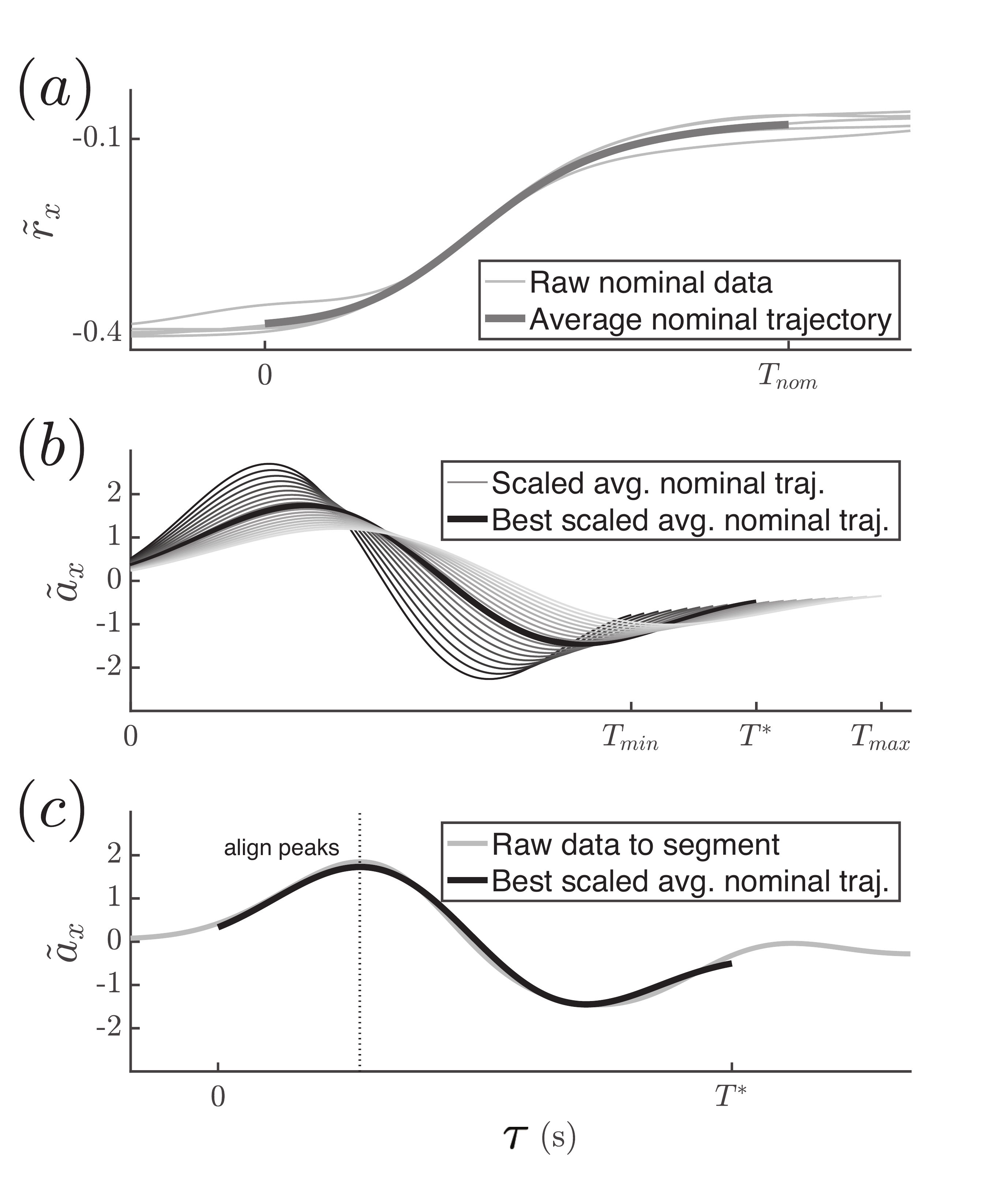}
	\caption{This figure illustrates the segmentation procedure.
	Step 2(b) of the procedure is depicted in Fig. \ref{fig:segmentation}(a).
	We take the subject's five nominal trials, shown in light grey, and average them to form a single average nominal trajectory, shown as the thick darker grey line.
	We choose a start and end time $[0, T_\regtext{nom}]$ for the average nominal trajectory as in Step 2(d).
	Step 3(a-b) of the procedure is depicted in Fig. \ref{fig:segmentation}(b).
	We obtain the horizontal acceleration trajectory of the average nominal trajectory by twice differentiating the position trajectory shown in Fig. \ref{fig:segmentation}(a).
	We then iterate over time scales in $[T_\regtext{min}, T_\regtext{max}]$, scaling the trajectory according to Sec. \ref{sec:methods:normalizing_trajectories}.
	These scaled trajectories are shown in shades of grey.
	An example trajectory corresponding to time $T^*$ is shown as the thick black line.
	Step 3(c-e) of the procedure is depicted in Fig. \ref{fig:segmentation}(c).
	The horizontal acceleration trajectory of a trial to be segmented is shown as the thin light grey line.
	Of the scaled trajectories in Fig. \ref{fig:segmentation}(b), the one corresponding to time $T^*$, shown as the thick black line, matches the best.
	The start and end times of the trial to be segmented are then determined by the endpoints of the overlaid black line.}
	\label{fig:segmentation}
\end{figure}
To segment the continuous kinematic STS data into individual trials, an appropriate start and end time for each trial must be chosen.
A consistent segmentation method is imperative for making comparisons across trials.
Ideally, the start and end of each trial is chosen so that important kinematic features of STS are aligned.

We developed the following segmentation procedure, displayed in Fig. \ref{fig:segmentation}, which is designed to align the peak horizontal accelerations of STS trials while being robust to differences in trial duration.
Portions of the procedure utilize a root-mean-square error (RMSE), defined as:
\begin{equation}
    RMSE(\hat{y}, y) = \sqrt{\frac{\sum_{i=1}^N (\hat{y}_i - y_i)^2}{N}}
\end{equation}
where $\hat{y} = \{ \hat{y}_i \}_{i=1}^N, \hat{y_i} \in \R$ and $y = \{ y_i \}_{i=1}^N, y_i \in \R$ are two sequences defined by sampling two different trajectories at $N$ times.

The procedure generates an average nominal trial from the subject's five nominal trials, and then uses it as a template to segment the rest of the trials.
We segment the natural and MT strategies' trials using the following procedure, but use a separate procedure for QS trials, detailed later on in this section.
\begin{enumerate}
    \item Manually segment each trial, starting several seconds before the subject begins to stand and ending several seconds after they have reached standing.
    \item To create an average nominal trial:
    \begin{enumerate}
        \item Take the manual segmentations of the subject's five nominal trials, and align their COM position trajectories such that the times at which the peak horizontal COM velocities occur are coincident.
        \item Average the subject's five nominal trials together to form a single average nominal COM position trajectory.
        \item Numerically differentiate the average nominal COM position trajectories to obtain velocity and acceleration trajectories.
        \item Choose the start time ($\tau = 0$) of the average nominal trajectory as when the horizontal COM acceleration first exceeds $20\%$ of its max. Choose the end time ($\tau = T_\regtext{nom}$) as when the vertical COM position first exceeds $99\%$ of its max.
        \item The average nominal trajectory is now defined over $[0, T_\regtext{nom}]$. 
        Follow the scaling laws in Sec. \ref{sec:methods:normalizing_trajectories} so that it is defined over $[0, 1]$.
    \end{enumerate}
    \item To segment each trial:
    \begin{enumerate}
        \item Iterate over times $T \in [T_\regtext{min}, T_\regtext{max}]$, where the interval is discretized into 100 samples. Let $T_\regtext{min} = 0.75s$ and $T_\regtext{max} = 1.5T_\regtext{nom}$.
        \item Rescale the average nominal trajectory to be defined over $[0, T]$.
        \item
        Align the scaled average nominal trial and the current trial so that the time at which the peak horizontal COM accelerations occur are coincident.
        \item The trajectories overlap in a window of length $T$ seconds. 
        Compute the RMSE between the horizontal COM acceleration trajectories over this window.
        \item Find $T$ that minimizes this RMSE (to accurately segment CP trials despite the effects of perturbation, we only minimize the pre-perturbation RMSE for CP trials).
    \end{enumerate}
\end{enumerate}

Since the QS strategy utilizes minimal forward momentum, it lacks a distinct peak in its horizontal COM acceleration trajectories when compared to the natural and MT strategies.
This is shown in Fig. \ref{fig:strategy_ax}.
Therefore, we developed a separate segmentation procedure for the QS strategy, where Steps 1 and 2 remain the same as before:
\begin{enumerate}
    \setcounter{enumi}{2}
    \item{To segment each QS trial:}
    \begin{enumerate}
        \item
        Iterate over times $T \in [T_\regtext{min}, T_\regtext{max}]$, where the interval is discretized into 100 samples. Let $T_\regtext{min} = 0.75s$ and $T_\regtext{max} = 1.5T_\regtext{nom}$.
        \item Rescale the average nominal trajectory to be defined over $[0, T]$.
        \item Manually align the scaled average nominal trial and the current trial so that the time at which the peak vertical COM velocities occur are coincident.
        \item The trajectories overlap in a window of length $T$ seconds.
        Manually choose $T$ that minimizes the RMSE between the scaled average nominal vertical COM velocity trajectory and the current trial's COM velocity trajectory. (Prioritize the pre-perturbation difference for CP trials, to accurately segment CP trials despite the effects of perturbation.)
    \end{enumerate}
\end{enumerate}
% \onecolumn
% \section*{Word Counts}
% This section is \textit{not} included in the word count. Also, it doesn't appear to be counting the abstract.
% % Don't count these!
% \quickwordcount{main}
% \quickcharcount{main}
% \detailtexcount{main}

%%TC:endignore

\end{document}